\documentclass[aps,prl,superscriptaddress,twocolumn,showpacs]{revtex4-1}
\usepackage{amsmath,amssymb,graphicx,epsfig,color,bbm,lipsum}
\usepackage[pass]{geometry}
\usepackage[colorlinks=true,citecolor=blue,linkcolor=blue,urlcolor=red,anchorcolor=black,filecolor=black]{hyperref}

\newcommand{\ba}{\begin{eqnarray}}
\newcommand{\ea}{\end{eqnarray}}

\begin{document}

\title{Evolving generalists via dynamic sculpting of rugged landscapes}

\author{Shenshen Wang}
\email{shenshen@physics.ucla.edu}
\affiliation{Department of Physics and Astronomy, University of California, Los Angeles, Los Angeles, CA 90095, USA}
\author{Lei Dai}
\affiliation{Institute of Synthetic Biology, Shenzhen Institutes of Advanced Technology, Chinese Academy of Sciences, Shenzhen 518055, China}


\begin{abstract}
Evolving systems, be it an antibody repertoire in the face of mutating pathogens or a microbial population exposed to varied antibiotics, constantly search for adaptive solutions in time-varying fitness landscapes. Generalists correspond to genotypes that remain fit across diverse selective pressures; cross-reactive antibodies are much wanted but rare, while multi-drug resistant microbes are undesired yet prevalent. However, little is known about under what conditions such solutions with a high capacity to adapt would be efficiently discovered by evolution, as environmental changes alter the relative fitness and accessibility of neighboring genotypes. In addition, can epistasis --- the source of landscape ruggedness and path constraints --- play a different role, if the environments are correlated in time? We present a generative model to estimate the propensity of evolving generalists in rugged landscapes that are tunably related and cycling relatively slowly. We find that environment cycling can substantially facilitate the search for fit generalists by dynamically enlarging their effective basins of attraction. Importantly, these high performers are most likely to emerge at an intermediate level of both ruggedness and environmental relatedness, trading diversity for fitness and accessibility. Our work provides a conceptual framework to study evolution in correlated varying complex environments, and offers statistical understanding that suggests general strategies for speeding up the generation of broadly neutralizing antibodies or preventing microbes from evolving multi-drug resistance.
\end{abstract}



\maketitle

\section{Introduction}


Temporally varying environments profoundly influence various properties of evolving systems, including their structure~\cite{hartwell:99, lipson:02, kashtan:05}, robustness~\cite{sasaki:97, thompson:00, csete:02, kitano:04}, evolvability~\cite{wagner:96, earl:04}, as well as evolutionary speed~\cite{kashtan:07} and reversibility~\cite{tan:12}.
Biological populations respond to environmental variations to minimize potential adverse effect on their survival and reproductive growth.
Adaptive solutions employed fall into two broad categories: generalists that perform reasonably well across environments, and a diverse mixture of specialists each excelling in a different environment.
Which solution confers the greatest selective advantage in the long run depends on the nature and statistics of environmental variations~\cite{shoval:12, mayer:17}.



Theoretical studies have examined the adaptive utility of survival strategies at different timescales of environmental fluctuations~\cite{kussell:05, kussell:06, mustonen:09, rivoire:11, mayer:16, skanata:16}. While stochastic switching between distinct specialist phenotypes appears to be favored when environments change sufficiently slowly~\cite{kussell:05}, adopting a single generalist phenotype is shown to be advantageous for rapid fluctuations (e.g. faster than cell division)~\cite{gaal:10, patra:15}. Notably, these studies often assume the environments to be unrelated, randomly fluctuating and having only few phenotypic dimensions. However, natural environments may well be partially related over the course of the system's adaptation. Furthermore, the high-dimensional evolutionary landscapes, a nonlinear mapping from genotype to function, ultimately guide the adaptive search in the sequence space.
Deep mutational scans~\cite{fowler:14} have mapped out modest-size functional landscapes in fine details for a variety of evolving systems including protein binding affinity~\cite{adams:16, olson:14} as well as viral growth~\cite{haddox:18} and infectivity~\cite{wu:14}, highlighting the significant role of epistasis---interaction between mutations---in sculpting landscape ruggedness and shaping viable paths of adaptation.
Yet, how these intra-landscape structures interplay with inter-landscape correlations to constrain or open pathways toward generalists is not understood.






Generalists can reuse partial solutions in new contexts and hence rapidly adapt to previously unseen environmental conditions. 
In other words, such evolvable solutions are capable of extracting common features from correlated environments.
From a landscape perspective, generalist \emph{genotypes} can be recognized as local fitness optima shared by distinct landscapes representing varied environments.
Inspiringly complementary examples of generalists in adaptive evolution present outstanding challenges and opportunities:
A celebrated instance is the discovery of broadly neutralizing antibodies~\cite{burton:12} that target relatively conserved features of fast evolving pathogens such as HIV and influenza, which can evade recognition by specific antibodies while remaining fit;
an undesirable circumstance is the emergence of multi-drug resistant bacteria~\cite{kim:14} and viruses~\cite{rhee:10}.
Attempts to elicit broad antibody responses and to prevent multi-drug resistance have thus far met with mixed success~\cite{klein:13, burton:16, haynes:16, toprak:12, schenk:14}, which calls for a better and unified understanding of how evolution discovers generalists in correlated and changing fitness landscapes (or seascapes~\cite{mustonen:09}).


Here we present a general theoretical framework to address the propensity of evolving generalists in high-dimensional environments that are \emph{tunably related} and cycling relatively slowly (Fig.~\ref{ls_switching}).
This is motivated by evolution of the adaptive immune system against natural pathogens or man-made antigenic stimuli (e.g. vaccines) that change slowly or are sampled sparsely over time, so that considerable immune adaptation occurs in each epoch.
Of particular interest are two related questions: (1) Whether and under what conditions can environment cycling grant long-term selective advantage to generalists? (2) How do epistasis and environmental relatedness together impact the diversity and accessibility of generalist genotypes?




By constructing and characterizing rugged landscapes with tunable correlations within and between them --- a distinctive feature of this work --- we find that environment cycling can substantially enhance the likelihood of evolving fit generalists compared with evolution in a constant environment.
Large enhancement requires sufficient, yet not overly strong, similarity in landscape topography; the key is that optimal environmental relatedness (e.g. favorable sequence overlap between vaccine components) should balance a tradeoff between the prevalence, fitness and accessibility of generalists, so that cycling can preferentially enlarge the attractor size of fit ones. We show in a phase diagram that such balance shifts with the amount of epistasis.
This suggests that we may exploit the fitness correlations within and across cycling landscapes to favor the emergence and expansion of fit generalists in a population, against the natural tendency toward evolving specialists in slowly varying environments.

\section{Model}
\subsection{Construction of tunably related rugged landscapes}
The landscape framework has been used to study physical properties of disordered systems (e.g. macromolecules~\cite{bryngelson:95}, glasses and spin glasses~\cite{frauenfelder:91, sastry:98}) as well as nonphysical phenomena ranging from biological evolution~\cite{kauffman:92} to neural computation~\cite{hopfield:82} and business management~\cite{rivkin:03}.
The unifying attribute of this framework is its statistical characterization of the global topography of a complex mapping.
Interest in fitness landscapes stems from the need for intuition into the evolutionary behavior of populations in the presence of epistasis~\cite{weinreich:06, poelwijk:07, kryazhimskiy:09, gong:13, weinreich:13, devisser:14, wu:16, macken:89, hwang:18}.
Epistatic interactions can result in mutations that are individually deleterious but jointly beneficial, giving rise to multiple local optima in a genotypic fitness landscape that represent degenerate solutions to a particular task.
Epistasis is central to understanding the predictability of evolutionary paths~\cite{kryazhimskiy:14, lassig:17} as well as evolvability~\cite{bloom:06, bloom:10, boyer:16} and adaptation rate~\cite{chou:11, kryazhimskiy:14} of biomolecules.

To determine general properties that arise solely from the global topography of landscapes (i.e., the degree and statistical structure of ruggedness), irrespective of the specific structure of the evolving system \emph{per se}, we use the NK model~\cite{kauffman:87} to represent generic rugged landscapes.
This paradigmatic family of model landscapes for protein evolution, inherently related to the spin glass model~\cite{SK:1975} in statistical physics, has yielded much insight into affinity maturation of antibodies~\cite{kauffman:89, perelson:95}.

In an NK landscape, the fitness of a genotype represented by a bit string $\vec{S}$ of length $N$ is defined as the average over each bit's fitness contribution:
\begin{equation}
\label{fitness}
f^{\epsilon}(\vec{S})=\frac{1}{N}\sum_{i=1}^Nf_i^{\epsilon}(S_i, S_{i_{1}},\cdots, S_{i_{K}}).
\end{equation}
Here the fitness contribution of bit $i$, $f_i^\epsilon$, in a given environment $\epsilon$ depends on $\{S_i, S_{i_{1}},\cdots, S_{i_{K}}\}\equiv\{S\}_i$, the state of the $K+1$ coupled sites influencing the fitness contribution of site $i$.
An additive landscape ($K=0$) has a single global optimum reachable from an arbitrary starting genotype, whereas in a completely random landscape ($K=N-1$) statistical independence of nearby states leads to an extraordinarily rough surface in which on average $2^N/(N+1)$ local maxima can be surrounded by deep valleys. Natural populations are likely to be guided by fitness landscapes in between these extremes.

In the case of antibody-antigen binding affinity, each distinct antigen defines a unique hypersurface spanning over a hypercube of $2^N$ binary antibody genotypes.
To link the level of fitness conservation (the likelihood that a fitness contribution is preserved across environments) to topographical relatedness between landscapes,
we consider two environments $A$ and $B$ in which
\begin{equation}
\label{ls_pair}
f_i^B(\{S\}_i)=a_i f_i^A(\{S\}_i).
\end{equation}
This constructs landscape $B$ from landscape $A$; the latter is generated according to Eq.~(\ref{fitness}) with $f_i^A\sim \mathcal{U}(-0.5,0.5)$ and randomly chosen interacting neighbors. Although more structured interaction schemes (e.g. block neighborhood) tend to modestly increase ruggedness~\cite{hwang:18}, this factor has little effect on our qualitative results.
The strength $a_i$ of correlation between fitness contributions of site $i$ in two environments is given by
\begin{equation}
\label{ai}
a_i=
\begin{cases}
1 & i\leq n_p \\-1 & i>n_p
\end{cases}
\end{equation}
This choice is supported by our analysis on the fitness landscapes of $\beta$-lactamase under different $\beta$-lactam antibiotics~\cite{mira:15} (see SI for details). We found that the fitness effect of single resistance mutations follows a bimodal distribution (i.e. with two peaks around $+1$ and $-1$, respectively), suggesting that in this empirical system the effect of single mutations under different environments is predominantly conserved ($a_i\sim1$) or subject to tradeoff ($a_i\sim-1$)(Fig.~S1 and Supporting Information Text).
While $n_p=N$ corresponds to identical landscapes, $n_p=0$ characterizes completely inverted pairs.
Therefore, the fraction of conserved fitness contributions, $n_p/N$, naturally measures the level of conservation.
Furthermore, by making $a_i$ independent of the state of the $K+1$ epistatically interacting sites, we assume that fitness correlations are preserved in all backgrounds, which decouples the effect of inter-landscape correlations (fitness conservation characterized by $n_p$) from that of intra-landscape correlations (epistasis measured by $K$).
This decoupling in turn implies that $n_p$ tunes the topographical similarity without affecting the degree of ruggedness. As a consequence, landscapes thus constructed are tunably related yet statistically equivalent (Fig.~S2); changing $n_p$ does not alter the expected number (panel A) and mean fitness (panel B) of local optima.



\subsection{Adaptive walks in cycling landscapes}
To focus on the effect of global landscape topography on evolutionary dynamics, we consider adaptive walks under strong selection and weak mutation. In this limit, an evolving population can be regarded as a point in the genotype space that moves along paths of increasing fitness in single mutational steps. The population size of interest is sufficiently large to suppress random genetic drift but not too large so that escape from local fitness optima in a static landscape is very unlikely~\cite{weinreich:05}. We further assume ``greedy hill climbing" by which any starting genotype can be uniquely associated with a particular fitness peak at the end of the walk; this algorithm thus divides the genotype space into gaplessly packed catchment basins each surrounding a local fitness optimum.

\begin{figure}[htb]
\begin{center}
\includegraphics[angle=0, width=0.95\columnwidth]{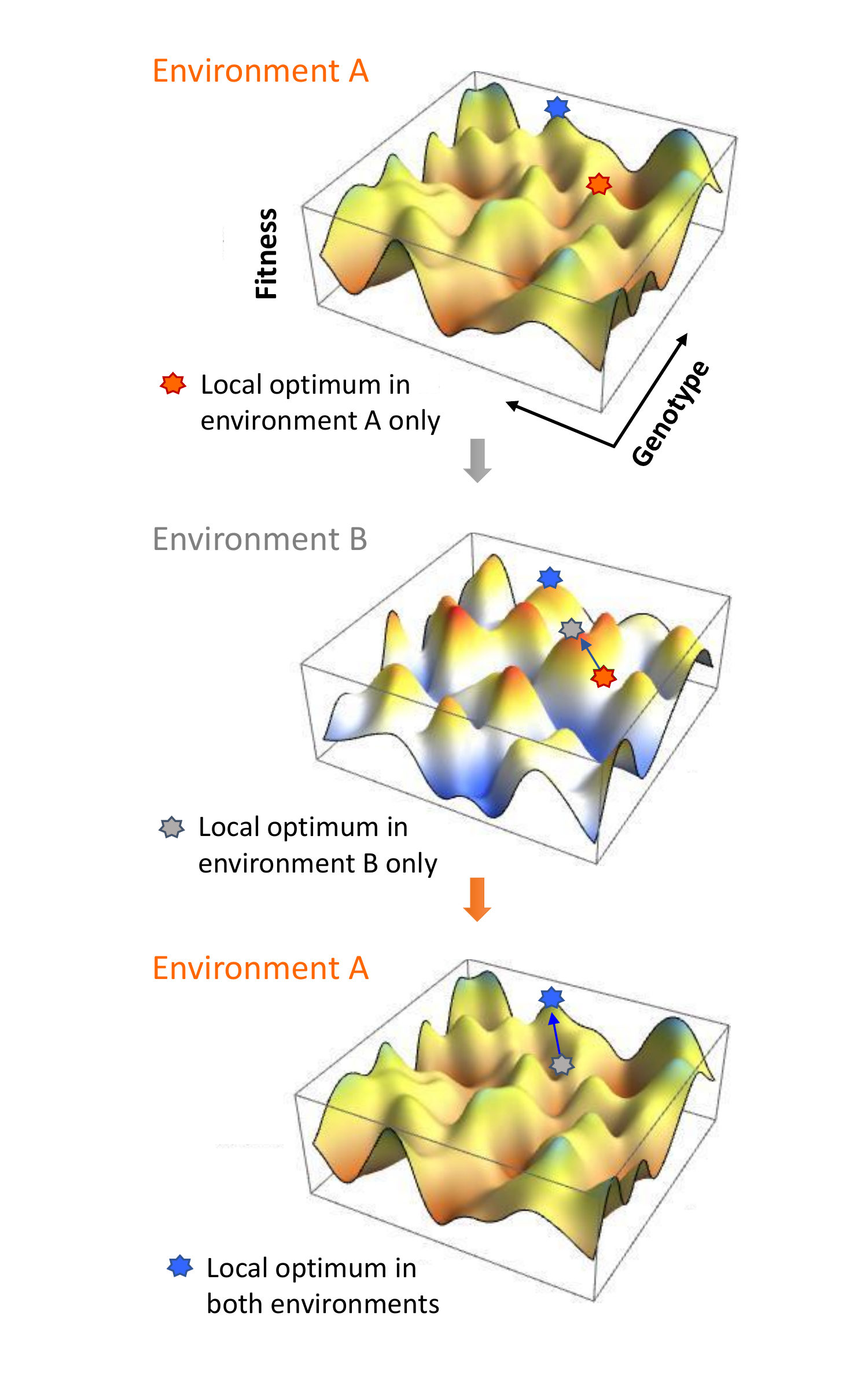}
\caption{A schematic view of cycling landscapes and an adaptive walk toward a generalist peak. Alternation between distinct yet correlated rugged landscapes can drive escape from fitness optima specific to one environment (orange/gray star in environment A/B) and open new paths (arrows) leading to genotypes locally or globally optimal in both environments (blue star). These shared peaks across landscapes --- generalist genotypes --- represent fixed points of adaptation in changing environments.}
\label{ls_switching}
\end{center}
\end{figure}

In the landscape description (schematic in Fig.~\ref{ls_switching}), generalist solutions can be identified as local fitness optima common to multiple distinct landscapes, whereas specialist solutions correspond to fitness peaks present in a single landscape. Switching between environments opens new possibilities not available in individual landscapes; environment cycling can free the population from a specialist peak (red/grey star in environment A/B) and produce an effectively continuous positive slope on the alternating landscapes, thereby creating evolutionary trajectories (arrows) toward a generalist peak (blue star in both environments) otherwise inaccessible from a specialist ancestor. Generalists thus act as stationary attractors in changing environments, i.e., fixed points of evolutionary dynamics.
Environment cycling is sufficiently slow but not too slow, so that in between switches the population is able to reach a local optimum and unlikely to escape from it.
As such, instead of performing an exhaustive study of adaptive dynamics, we directly characterize constituent landscapes and quantify their relationship.








\section{Results}

Our generative model links fitness conservation to topographical similarity. Our task then boils down to identifying topographical features that characterize the extent of relatedness, in a way that these static characteristics can inform the prospects for evolving generalists including their prevalence, fitness and accessibility.
For concreteness, we set $N=12$ and vary $K$ and $n_p$. In all plots, data are averaged over 1000 pairs of landscapes.

\subsection{Optimum sharing}

Local fitness optima that remain in the same location as the environment changes---shared optima across landscapes---represent generalist solutions.
Intuitively, as the number $n_p$ of conserved fitness contributions increases, it is more likely that the immediate neighborhood of local peaks remains and hence a greater prevalence of generalists is expected (Fig.~S3A).
Note that the expected number of shared optima between landscapes with equal amounts of conserved and sign-flipped fitness contributions ($n_p=N/2$) is identical to that between independent landscapes.

\begin{figure}[htb]
\begin{center}
\includegraphics[angle=0, width=1\columnwidth]{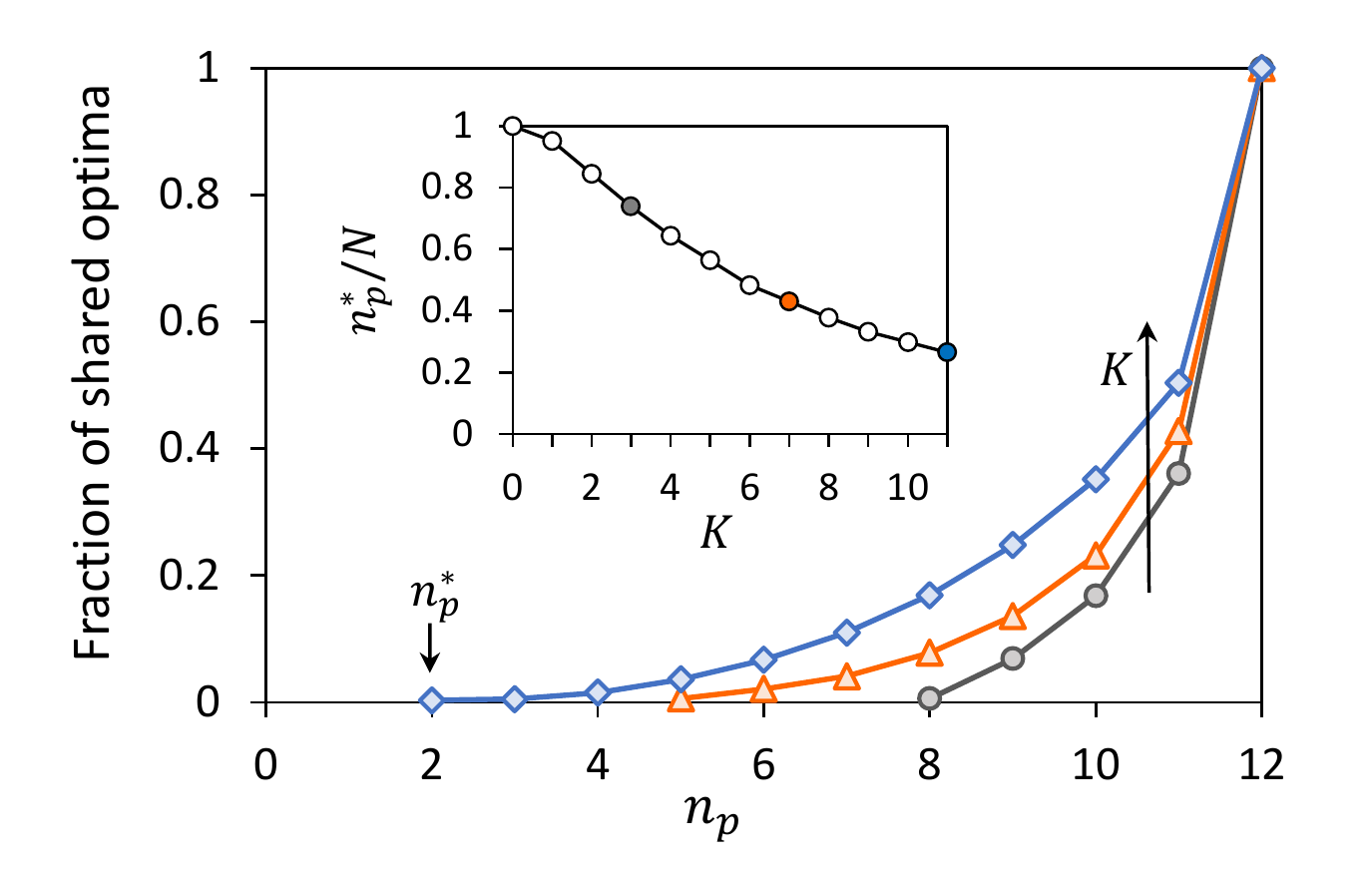}
\caption{The onset of optimum sharing and the abundance of shared optima. The fraction of local optima being shared between landscapes is shown as a function of the number of conserved fitness contributions ($n_p$). The minimum value of $n_p$ leading to optimum sharing is indicated with $n_p^*$. Curves correspond to $K=3, 7, 11$, increasing in the direction of the arrow. The inset shows $n_p^*/N$ as a function of $K$. Each data point is an average over 1000 landscape pairs.}
\label{opt_sharing}
\end{center}
\end{figure}

To exclude the effect due to the rapidly growing number of local fitness optima with the size $K$ of the epistatic groups, we plot the \emph{fraction} of local optima being shared between landscapes (Fig.~\ref{opt_sharing}) and observe two features as $n_p$ increases.
First, there is a minimum level of fitness conservation, $n_p^*/N$, below which no single generalist even exists; in this no-sharing regime, none of the genotypes remains locally optimal as the environment alters, i.e., all adapted states are specialists.
Second, both the onset of optimum sharing (at $n_p^*/N$) and the rate of growth in sharing depend on $K$. In particular, increasing $K$ weakens the dependence on $n_p$ of the degree of optimum sharing; as $n_p$ decreases, the fraction of shared optima decreases more slowly at larger $K$. The decline is nearly exponential at $K\simeq N/2$ and is faster (slower) than exponential for $K\leq(\geq)\,N/2$. Therefore, stronger ruggedness appears to promote optimum sharing, both by boosting the prevalence of generalists at a given conservation level (Fig.~2, $K$ increasing in the direction of the arrow) and by extending their presence to a lower level of fitness conservation (Fig.~2 inset).

This finding is somewhat surprising given that increasing ruggedness is often thought to imply reduced fitness correlations, until we realize that the impact of $K$ on landscape topography is not merely controlling the abundance of local optima, but also affecting how they are organized in the genotype space. When $K$ is small, the highest peaks tend to locate close to one another and the general configuration of the landscape is very non-random. As $K$ increases, fit local optima become more evenly distributed which may therefore raise the chance of peak sharing between landscapes.
An interesting implication thus follows: while increasing epistasis would reduce fitness correlations within a landscape, it might enhance correlations between landscapes at a given conservation level of fitness contributions.

When are generalists favored over specialists?
As known from ecology, generalist birds with intermediate bill lengths may evolve when prey types are alike, whereas specialization develops when more diverse prey types require highly dissimilar beaks. This also applies to the competitive advantage of generalist antibodies over specific ones in recognizing structurally related antigens. Our tunably related NK landscapes capture this trend (Fig.~S3B): The average fitness of shared optima increases with $n_p$ sublinearly; while in dissimilar environments (small $n_p$) specialists are on average more fit, at sufficiently high levels of environmental relatedness (large $n_p$), generalists become selectively favorable (arrows indicating the crossing between the average fitness of generalists alone, shown in solid lines, and that of specialists and generalists combined, shown in dashed lines).
Note that stronger ruggedness enlarges the generalist-favored regime toward a lower conservation level, at the expense of a modest reduction in average fitness.




\subsection{Dynamic basin linking}


\begin{figure*}[htb]
\begin{center}
\includegraphics[angle=0, width=1.8\columnwidth]{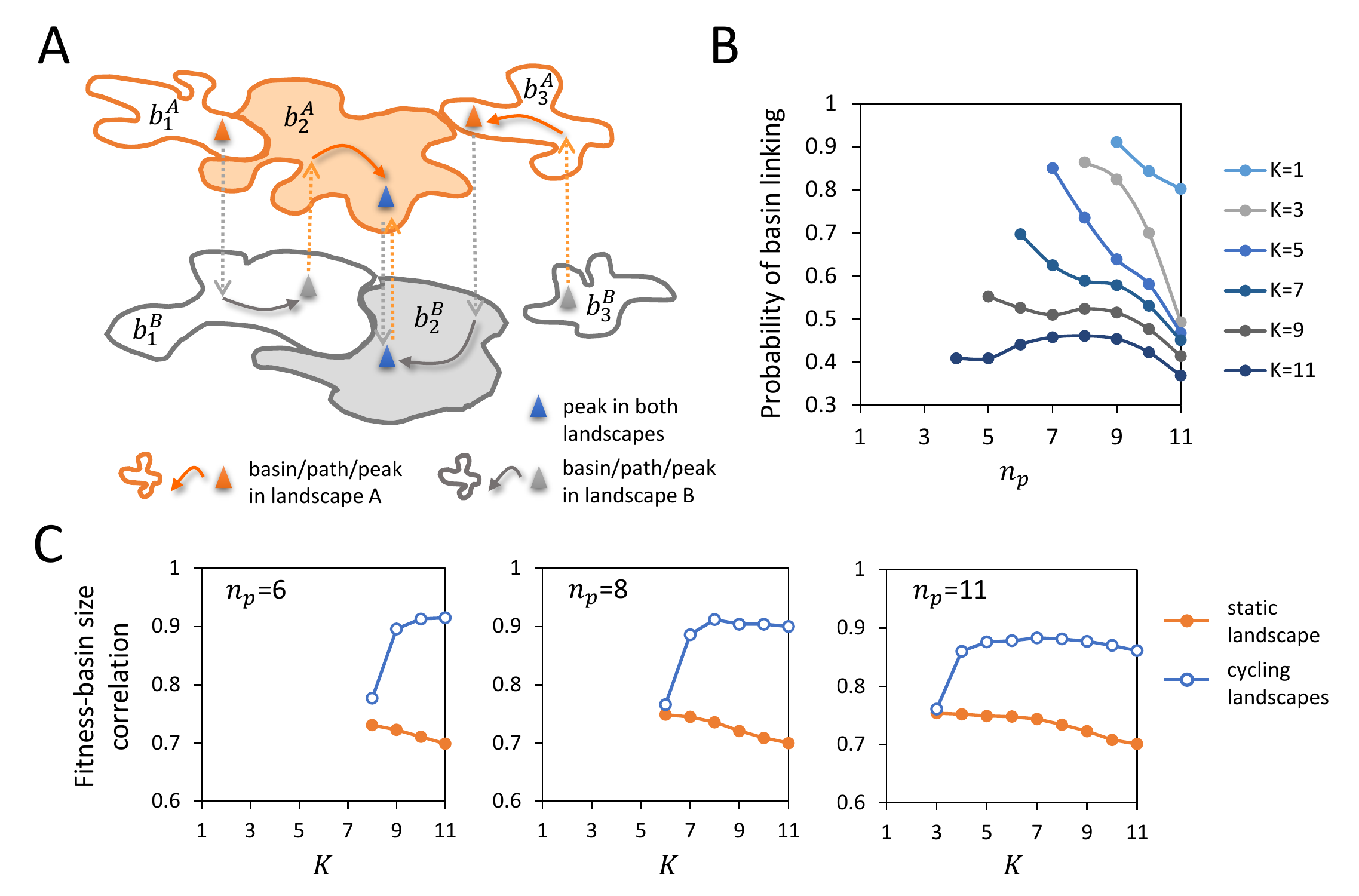}
\caption{Dynamic basin linking via landscape cycling. (A) Schematically, in a static landscape, only a single basin of attraction (shaded orange/gray in landscape A/B) leads to a generalist solution (blue peak). In contrast, under four rounds of landscape switching (ABAB or BABA, indicated with dotted arrows), all the genotypes located in the now linked basins (six amorphous shapes with orange or gray borders) can reach the generalist peak. This chain of basins $\{b_l^\epsilon\}$ has $b_2^A$ and $b_2^B$ that share the generalist peak serving as the hinge and the rest either as links ($b_1^B$, $b_3^A$) or as the ends ($b_1^A$, $b_3^B$). Therefore, the total coverage of linked basins defines the effective accessibility of a generalist in switching landscapes. (B) The fraction of generalists that gain basin size under landscape switching, where $K$ increases from top to bottom. Each data point is an average over $1000$ realizations of landscape pairs. (C) Correlation between the fitness and basin size of local optima in a static landscape (orange filled symbols) and in switching landscapes (blue open symbols).}
\label{bs_linking}
\end{center}
\end{figure*}

At a level of fitness conservation that supports optimum sharing between correlated landscapes, if the environment were static, only starting genotypes in a single basin of attraction (e.g., $b_2^A$ or $b_2^B$ in Fig.~\ref{bs_linking}A, the ramified shape filled with orange or gray color) would lead to the encompassed generalist peak (blue triangle).
In contrast, environmental alternation (dotted arrows) might link to the ``hinge" basins ($b_2^A$ and $b_2^B$) additional basins that surround specialist peaks (orange or gray triangles) and are otherwise disconnected in individual landscapes. Each successively linked peak is determined as the highest local optimum in current environment that is enclosed by a basin in the preceding environment (e.g., the peak in basin $b_3^A$ is the fittest genotype in landscape $A$ that belongs to basin $b_2^B$ in landscape $B$). In this way, landscape cycling dynamically enlarges the attractor size of generalists via connecting basins disjoint in static environments (e.g. $b_1^B$ serves as a bridging basin between $b_1^A$ and $b_2^A$);
such valley-bridging effect of environmental changes has been observed in engineered bacteria~\cite{steinberg:16}.

It is important to note that basin linking via landscape switching enhances the population flux from specialists to generalists but not much the reverse.
In other words, cycling between correlated environments creates a ratchet-like effect on the population flow, driving it away from specialists that see large fitness swings as the environment alters and toward generalists that experience little fluctuations in selection pressures.
Consequently, starting from any genotype located inside these dynamically linked basins, the population would converge to the generalist peak after a sufficient number of environmental switches ($ABAB$ or $BABA$ in the example in Fig.~\ref{bs_linking}A).
Therefore, the total coverage of linked basins defines the effective accessibility of a generalist in cycling environments.
Note that the number of basins in a chain is modest (e.g. no greater than $8$ for $N=12$; see Fig.~S4, panels A and C). Thus, for landscape sizes relevant to antigen or antibiotic binding sites, several environmental cycles would suffice to channel the population to generalists.

We next quantify potential benefit of landscape cycling.
We first estimate the probability that generalists have greater accessibility under environment cycling than in a static environment. Specifically, we evaluated the fraction of shared optima that acquire additional basin size via basin linking (Fig.~\ref{bs_linking}B).
At high levels of fitness conservation (large $n_p$), the frequency of basin linking declines with increasing $n_p$; too similar landscape topography makes it unlikely that a generalist-encompassing basin in one landscape contains a specialist peak in the other landscape. Increasing epistasis also monotonically diminishes the chance of basin linking; stronger decorrelation in fitness (larger $K$) results in fewer (Fig.~S4C) and smaller (Fig.~S4D) linked basins.
In highly random landscapes ($K\geq 9$), the likelihood of basin linking exhibits a relatively weak dependence on $n_p$, yet showing a maximum at intermediate values of $n_p$. This corresponds to, on average, longer chains (Fig.~S4A) of larger basins (Fig.~S4B) compared to weaker or stronger fitness conservation, reflecting an optimal complementarity between landscape profiles that supports optimum sharing without over-suppressing basin linking.

To explore how landscape cycling might affect the difficulty in evolutionary search for fit generalists, we computed the correlation coefficient between the fitness and basin size of shared optima (Fig.~\ref{bs_linking}C). For a single static landscape, the correlations are already positive and high, decreasing with increasing epistasis, which is consistent with known properties of NK models. Remarkably, under landscape cycling, total basin size of linked optima and their overall fitness are much more correlated compared to the static case; such enhancement in fitness-basin size correlation is significant as long as optimum sharing is prevalent. The strongest enhancement again occurs at intermediate values of $n_p$, showing little decline toward larger $K$.
Taken together, landscape cycling can significantly enlarge the catchment basins of generalists, especially for those with high fitness, well beyond the counterpart in static environments.



\subsection{Likelihood of evolving fit generalists}

\begin{figure*}[htb]
\begin{center}
\includegraphics[angle=0, width=1.8\columnwidth]{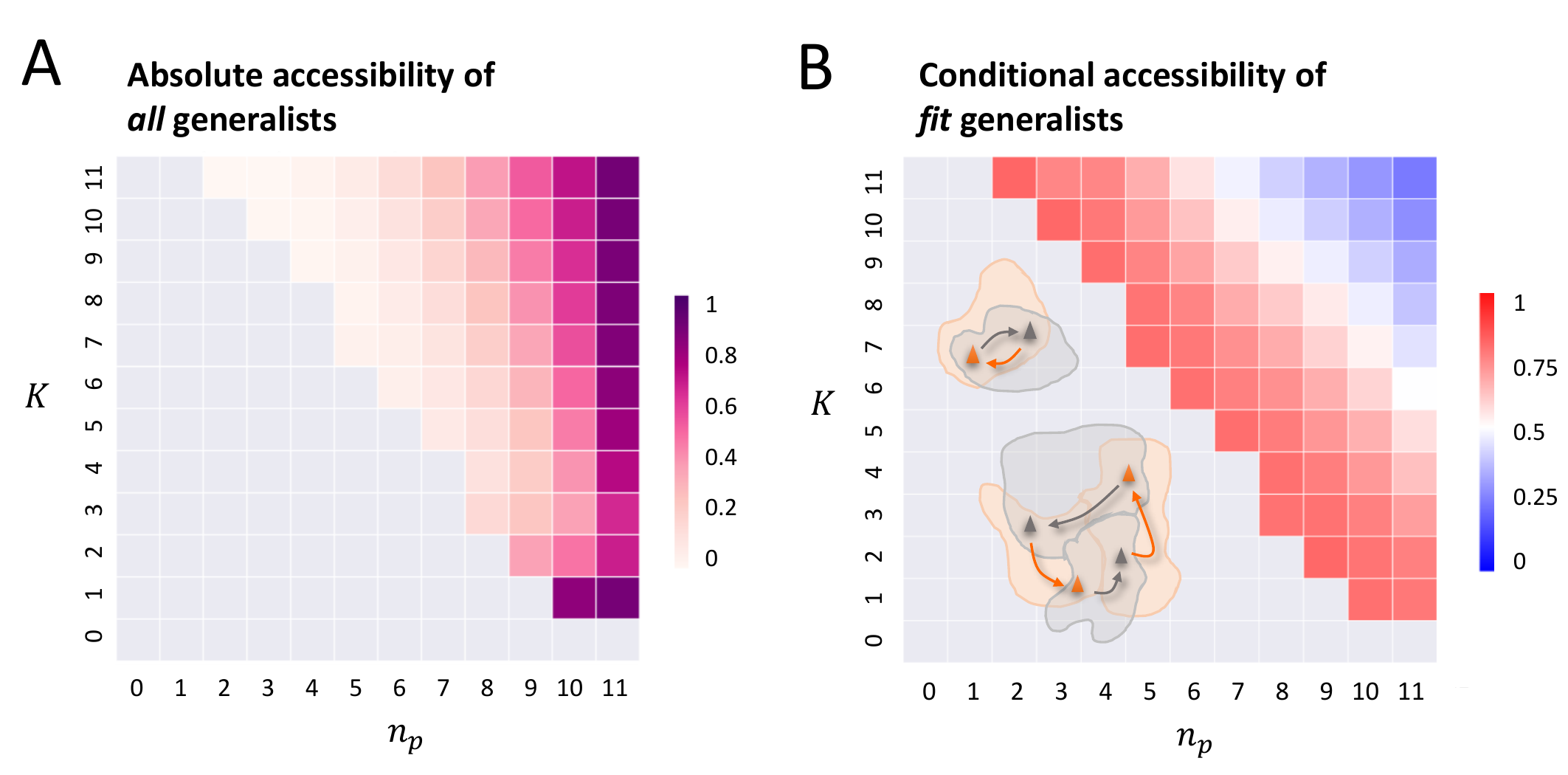}
\caption{Accessibility of generalists under landscape cycling. (A) The fraction of genotypes that can reach a shared optimum via an adaptive path. (B) The ratio of the total basin size of fit shared optima (within the top $30\%$ of maximum fitness) to that of all shared optima. Both heatmaps are obtained by averaging over 1000 pairs of landscapes at each combination of $n_p$ and $K$. In both diagrams, the gray area corresponds to phase I in which all local optima are specific to one environment; in this no-generalist phase, landscape switching leads to either oscillations between two specialist peaks each in one landscape (panel B, upper inset) or limit cycles (lower inset).
The colored region represents phase II, a convergence phase, where generalist peaks act like hubs into which evolutionary trajectories enclosed by the corresponding linked basins converge, following multiple landscape switches.
}
\label{access}
\end{center}
\end{figure*}

At similar levels of epistasis, landscape topography can nevertheless differ markedly. Whether generalists would benefit from landscape switching depends critically on the topographical compatibility between alternating landscapes. As shown in the phase diagrams (Fig.~\ref{access}), for a given $K$, as $n_p$ increases, the system crosses the boundary from phase I in which all adapted states are specialists (grey region) to phase II where generalists constitute stationary attractors in alternating landscapes (colored region).
While selectively accessible (i.e. monotonically increasing in fitness) paths are rarely circular on a static landscape, closed paths may prevail under environmental cycling in phase I --- either due to oscillations between nearby peaks each present in only one landscape (Fig.~\ref{access}B upper inset), or arising from limit cycles composed of specialist peaks located in successive basins on alternating landscapes (Fig.~\ref{access}B lower inset). In both scenarios, the population is pushed away from a local optimum upon every switch and never settles. In contrast, in phase II, landscape cycling can drive a population flow to a generalist peak.
Note that even in phase II specialists and generalists would coexist, because initially-specialists would either remain specialists if they start in isolated basins, or they are in transit via linked basins toward a generalist peak. This is relevant to the composition of immune repertoires, as instantaneous products (e.g. antibodies) accumulate throughout the entire course of immune response.

To estimate the mutational accessibility of generalists in slowly cycling environments (phase II), we first computed the fraction of genotypes that would follow an adaptive path to a shared optimum as landscapes alternate (Fig.~\ref{access}A).
This quantity, which measures the total accessibility of all generalists combined, is large either when the number of shared optima is large at large $n_p$ and large $K$ (Fig.~S5A) or when the basin size of shared optima is large at small $K$ (Fig.~S5C). Yet, among these generalists-to-be, the accessibility of the fit ones (defined as being within the top $30\%$ of the maximum fitness among the shared optima) determines the likelihood of evolving fit generalists. Fig.~\ref{access}B shows a heat map of the expected ratio of the total basin size of fit shared optima to that of all shared optima for each combination of $n_p$ and $K$. This conditional accessibility decreases monotonically as $n_p$ and/or $K$ increases, once exceeding the onset of optimum sharing (i.e. in phase II). Notably, the chance of evolving fit generalists is worst in the strong-conservation high-epistasis corner (blue color), where the sequence space divides into many small and rarely linked basins surrounding shared optima of which the majority are unfit (Fig.~S5A and Fig.~S5B, while the total number of generalists rapidly grows with increasing $n_p$ and $K$, the fit ones saturate in number).

Therefore, to enable efficient discovery of fit generalists, cycling rugged landscapes should have an adequate level of epistasis to allow a diversity of solutions (Fig.~\ref{opt_sharing}), while presenting complementary profiles of ruggedness to guide adaptation, enlarging the set of mutational trajectories leading to fit shared optima (Fig.~\ref{bs_linking}). Interestingly, the conserved fraction of sitewise fitness contributions---a mean-field-like parameter---closely tunes the topographical correlations between landscapes: an intermediate level of conservation encourages adaptive linkage of successive specialist basins in alternating landscapes toward the generalist peak. Such cycling induced basin linkage preferentially enhances the accessibility of fit generalists (Fig.~\ref{access}B) over less fit ones.
This behavior is statistically robust and represents a balance of tradeoff between the abundance, quality and searchability of evolvable solutions in a high-dimensional nonlinear map from sequence to function.

\section{Conclusion and discussion}

We present an attempt to endow the ecological notion of generalists with an evolutionary meaning in the context of adaptive strategies in evolving systems. Specifically, we demonstrate the impact that cycling between distinct yet related environments might have on the discovery of generalists---genotypes adapted to recurring features in changing environments.
We provide a statistical framework to construct and characterize tunably related fitness landscapes analogous to spin glass models, and extend the idea of adaptive walks to study long-term evolution in environments that change on comparable timescales as population adaptation.
We show that landscape topography and relatedness interplay to determine the relative prevalence and fitness of specialist and generalist genotypes.
Depending on the degree of fitness conservation and the amount of epistasis, evolutionary dynamics divides into two phases: (I) oscillations or cycles among specialist peaks in the absence of generalist solutions, and (II) convergence to a generalist peak after multiple environmental switches.
We find that an intermediate amount of epistasis, reflective of evolved functional constraints in biological systems, appears to balance the abundance (Fig.~\ref{opt_sharing}) and accessibility (Fig.~\ref{bs_linking}B) of generalists. What is more, in the convergence phase, an intermediate level of similarity between the structure of ruggedness in alternating landscapes affords the best chance of evolving fit generalists (Fig.~\ref{access}), by more effectively enlarging their basins of attraction (Fig.~\ref{bs_linking}B) and strengthening the correlation between basin size and fitness (Fig.~\ref{bs_linking}C) compared to weaker or stronger similarity.

In the context of adaptive immunity, one of the authors and coworkers have shown that temporal correlations in antigenic environments crucially regulate evolutionary pathways of lymphocyte populations---for one~\cite{wang:15} or multiple~\cite{wang:17} binding targets--- weighing evolvability against viability.
Here, we make this notion more precise: by describing the underlying rugged landscapes in a statistical manner, we turn the abstraction of environmental correlations into concrete measures of relatedness between landscapes---such as the frequency of optimum sharing and basin linking---and predict adaptive outcomes based on these topographical attributes.
These predictions are relevant because they rely on a statistical framework that neither oversimplifies the topography for interstate dynamics, nor fully characterizes all possible evolutionary trajectories which is not practical for system sizes of most interest.
This generic approach thus helps uncover key determinants of the propensity of evolving fit generalists, emphasizing the importance of \emph{simultaneously} considering the role of epistasis and topographical compatibility between landscapes in guiding the evolutionary discovery. To more efficiently induce broadly neutralizing antibodies, our findings suggest cycling between antigens that are similar enough to allow generalists to exist and yet sufficiently different to enlarge the basin of fit ones.

Although the present study is motivated by finding conditions that foster the evolution of generalists, it also reveals strategies to slow their emergence.
For the latter, as shown in the phase diagram (Fig.~\ref{access}), dissimilar landscapes (small $n_p$ and modest $K$) could avoid generalists and confine the population to a small region of the sequence space under landscape switching, whereas very similar rugged landscapes (large $n_p$ and large $K$) may trap the population to unfit generalist genotypes. These results further our understanding of multi-drug resistance by unifying diverse studies in a common framework. For example, in the case of antibiotics, experiment has shown that alternating environments can constrain the evolution of multi-drug resistance, in the regime where $n_p$ is small between different antibiotics~\cite{kim:14}. By contrast, for HIV protease inhibitors, $n_p$ is relatively large and hence multi-drug resistance is easily achieved~\cite{rhee:10}.
Thus, our results provide a guide for choosing drug combinations.

Our model is readily extendable to study other aspects of adaptation in changing environments within a landscape framework. First, the valuable information provided by deep mutational scans can be used to extract inter-landscape correlation strength for different orders of epistatic interactions. Our method can then map out the phase diagram which informs what parameter regime holds the best promise for desired evolutionary outcomes.
Second, our current predictions are intended for long-term adaptation of large populations under relatively slow environmental alternation. In future work, we will study the impact of switching rate on evolutionary dynamics of finite populations and determine when the predictions might depart from the asymptotic limit.
In addition, the measures of relatedness can be generalized to account for cycling among a larger number of environments.
Finally, many of our results based on temporally varying environments can be extended to understand adaptation in spatially heterogeneous environments, where landscape switching arises from migration between distinct yet connected habitats or microenvironments~\cite{zhang:11}.












\section{Acknowledgements}
S.W. gratefully acknowledges funding from UCLA.
L.D. was supported by Jane Coffin Childs postdoctoral fellowship.

\bibliography{Ref}



\end{document}